# An Optimized Microeconomic Modeling System for Analyzing Industrial Externalities in Non-OECD Countries

Agnibho Roy[1], Abhishek Mohan[2]

[1]*Coppell High School, Coppell, Texas, 75019, United States*
[2]*Texas Academy of Mathematics and Science, Denton, Texas, 76201, United States*

**Abstract:** In this paper, we provide an integrated systems modeling approach to analyzing global externalities from a microeconomic perspective. Various forms of policy (fiscal, monetary, etc.) have addressed flaws and market failures in models, but few have been able to successfully eliminate modern externalities that remain an environmental and human threat. We assess three primary global industries (pollution, agriculture, and energy) with respect to non-OECD entities through both qualitative and quantitative studies. By combining key mutual points of specific externalities present within each respective industry, we are able to propose an alternative and optimized solution to internalizing them via incentives and cooperative behavior rather than by traditional Pigouvian taxes and subsidies.

**Focus Points and Keywords:** Pollution, agricultural economies, global energy, OECD, systems modeling, externalities, microeconomics, taxes, and subsidies.

## 1. Introduction

The development of new businesses and industries has allowed for improved human productivity and efficiency of resources.[1] However, the repercussions of physical and chemical processes have developed into a primary concern for both OECD and non-OECD nations. Microeconomic analyses of theories and their applications to current societies prove to be the backbone in developing a thorough plan that curbs air pollution, energy crises, animal endangerment, etc., but previous models fail to account for the possibilities that emerge from technological innovation and cleaner energy production. Take for example the merging of superconductivity and TI research with microscope industries to develop SQUID microscopes. As a result, technology emerging from the cooperation of the two industries has resulted in utilizing the applications of quantum mechanics and dissipation less edge current, allowing for essentially perfect electricity reusability and efficacy.[2] Such behavior internalizes the externality of excess bi-product gases such as $CO_2$ and $CH_4$ released from natural gas derived electric power, returning production to the socially optimal level (consider minor adjustments be made by Pigouvian taxes or subsidies: government regulations that correct for various externalities).

In the aforementioned economic scenario, an externality refers to a consequence of an economic activity, being a cost or benefit, which affects a separate party that did not intend on incurring the resulting consequence. In essence, the social cost of the activity is sufficiently heavier than the private cost, but this effect can be eliminated by internalizing them to the point where the social cost is approximately equal to the private cost.[3] Externalities can most specifically be observed within the consumption and production activities of third-party groups. The OECD, or Organization for Economic Co-operation and Development, is an established international association that promotes global trade and intercommunication, in addition to universal economic and social well-being. OECD nations and their dynamic relationships with global externalities provide essential information for optimizing current microeconomic modeling systems and allowing other non-OECD nations to join the elite economic societies.[4] Industrial activities are a common method used in order to better understand and analyze externalities on an international level. Furthermore, the selection of the three specific externalities provided by pollution, agricultural economies, and global energy consumption are informative means of further optimizing current economic modeling methods.

Multiple varied modeling systems currently exist that provide insight into understanding the influence of externalities, both positive and negative, on global economic dynamics.[5] However, a model based on conceptualized and diagram-based systems modeling has not yet been entirely developed for specifically analyzing the previously discussed industrial externalities in conjunction with one another. Addressing this area of interest, this paper suggests a microeconomic method of analyses that explicates the nature of business behaviors and the share of burdens of the externalities they produce within their respective industries. Additionally, we combine the key mutual points of each externality and propose an alternative solution to internalizing them via incentives and cooperative behavior rather than traditional Pigouvian taxes and subsidies.

## 2. Pollution: Industrial Analysis

The OECD has previously provided a comprehensive assessment of the negative influences outdoor air pollution imposes upon the entirety of both its internal and external entities. After identifying the key factors in mortality and alterations in pollutant concentration of OECD and non-OECD countries, the rapidly expanding threat of harmful emissions, and the developing presence of global warming, we believe that air pollution requires more firm regulations. As various international organizations have understood the relevance of this issue, we aim to better understand and graphically represent the influence of pollution in non-OECD countries. Thus, we introduce our analysis on air pollution in terms of fine particulate matter ($PM_{2.5}$) as concentration per cubic meter (1 atm pressure).[6]

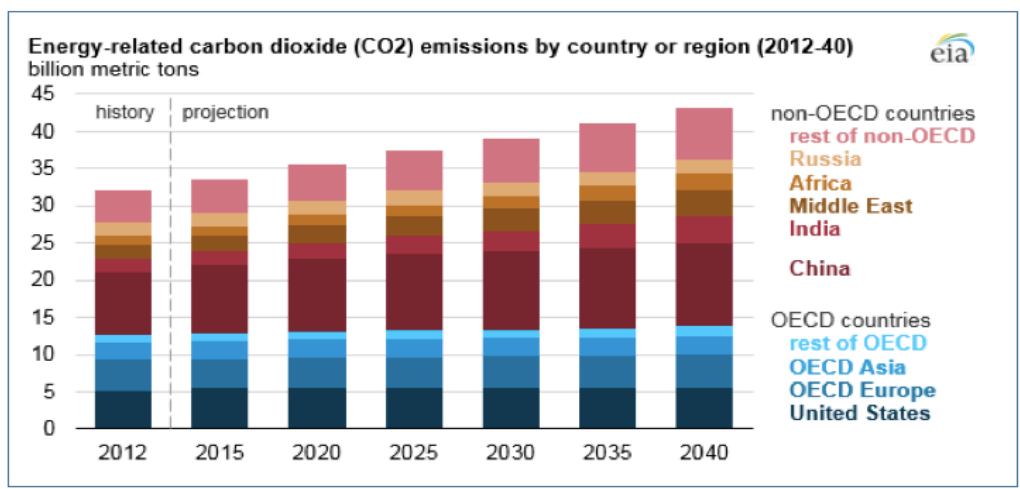

*Figure 1:* EIA provides a graph to depict the increase in the $CO_2$ emissions over time by extrapolating data from previous years, which is color coded by OECD and non-OECD nations in addition to sub-regions within each organization.[7]

Concentrations, as stated by the WHO, have depicted a clear trend between OECD and non-OECD nations in 2015 data: the average median fine particulate matter concentrations in non-OECD nations far exceed those of OECD nations.[8] In addition, one needs to define the elasticity of marginal social benefit (MSB), marginal private benefit (MPB), marginal social cost (MSC), and marginal private cost (MPC) relative to the level of pollution for a non-OECD nation in order to provide key information about surplus and deadweight loss.

Regarding the MPB, businesses benefit from being able to pollute as a bi-product of their innovations, as they lack the need to seek alternative emission-reducing methods; however, the benefit will decrease over time due to diminishing marginal utility. Additionally, it can be reasonable to assume that the MPB approaches zero as pollution increases since it will have no effect on business decisions at a certain threshold. MPC remains constant (or with a slope slightly higher than zero) due to some regulation effect harming them for their pollution emission.

However, the MSC would indeed be much steeper than the MPC as an additional unit of pollution would harm those in proximity to the pollution in comparison to simply harming the business. For the MSB, the business is part of society and the rest of society receives no additional benefit, and thus, the curves are equal.

The solution in proposal focuses primarily on the rate of change of the benefit to society from an additional unit of pollution, and such can only be accomplished through cooperative processes. Two essential partnerships may be emphasized: air purifying and regulatory practices and environment-friendly technology with traditional $PM_{2.5}$ emitting businesses and industries.

The hybridization of cars has allowed for the decrease in emissions from cars as well as improved efficiency. Data from Argonne National Laboratory has shown the decrease from 6500 Btu to 4200 Btu for a standard gasoline-powered and hybrid vehicle, respectively, which is in direct proportion to emissions released.[9]

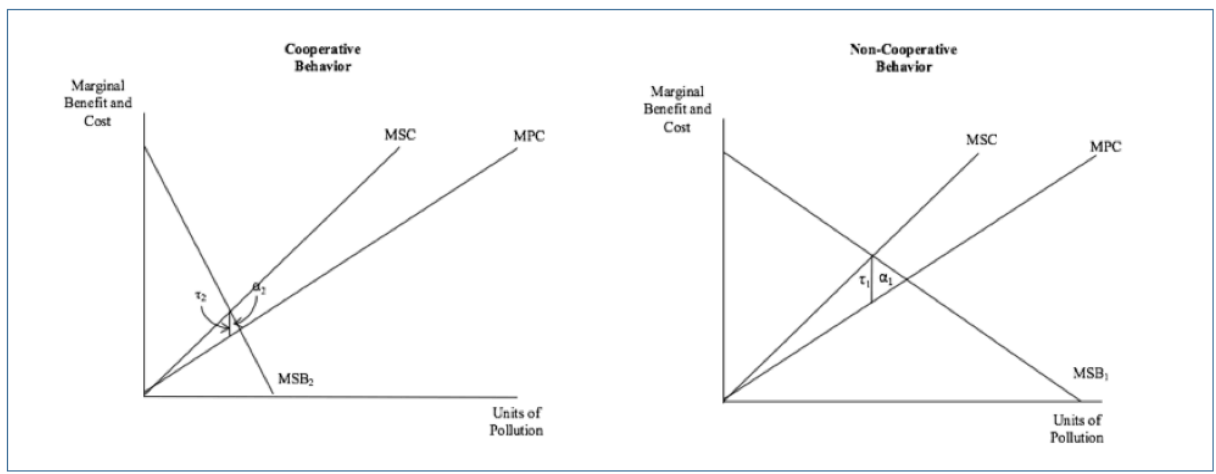

*Figure 2: A graph designed by the authors to visually display the theoretical behavior of economies in terms of marginal benefits and costs. Notice in cooperative behavior, the slope of the MSB is lower than the MSB in non-cooperative behavior.*

As a result, improved technology and transfer to hybrid transport systems in non-OECD nations have the potential to reduce $PM_{2.5}$ and $CO_2$ emissions by half per unit. Additionally, air regulatory controls such as the ICAC have developed additional controls and information aimed to reduce emissions and foster business growth.[10] Both factors would increase the magnitude of the slope of the MSB, as an extra unit of pollution would provide even lower benefit than before. A summary of these trends is depicted in Figure 2. Further mathematical analyses will be presented in the final section to display the improvement in the welfare economics as a result of this business cooperation. However, before doing so, other externalities shall be presented so that the final optimized model may be a solution to understanding a wide breadth of economic and environmental problems.

## 3. Agriculture: Industrial Analysis

Agriculture is an essential unit in a country's economic development and security, serving as a primary source of sustenance for its individual populations. For both OECD and non-OECD entities, agricultural production and productivity have seen drastic changes over the past few decades, primarily attributed to both technological innovation and enhanced global interconnection in trading and affairs. However, OECD countries have begun to represent a rapidly declining share of agricultural output on the international scale, with other such non-OECD countries encompassing a growing share.[11] The disparity between the two distinct groups provides great room in better understanding their individual influences upon global agricultural output and efficiency, and more specifically, the future adaptive market efficiencies on the global market. This understanding, in conjunction with non-OECD countries, provides an interesting and relevant area for further modeling optimization. Thus, we introduce our analysis on the agricultural industry by attempting to internalize the effect of agricultural output on the following externalities: water contamination, wildlife contamination, and natural aesthetics.

The initial externality to consider is water contamination as a result of agricultural processes. The US EPA has regarded agriculture as the leading source of water quality impairment in the United States since 1994, with the rate only increasing, making the externality of great concern in terms of MSC.[12] However, the key statistic to note is the nitrate contamination in the groundwater, a serious contaminant. Figure 4 represents the nitrate concentration as a function of time for the agricultural-based, non-OECD nation of Sri Lanka; the figure is specifically divided into the dry and rainy seasons. The fact that the contamination differs from a cultivated to a non-cultivated area by over 45 milligrams of N/Liter of water indicates the inefficiency and uncleanliness of agricultural processes of non-OECD nations. Furthermore, with the exception of 3-4 points per curve, the groundwater exceeds the WHO cap for the highest allowable amount of Nitrate, which is the majority of each calendar year. Such high levels pose a problem in society as it threatens wildlife, primarily sea life contamination, and may be a contributing factor to the current sixth mass extinction, where extinction is occurring 100 to 1000 times the natural extinction rate.

A presented scenario may also portray the obstruction of natural aesthetics in the community due to agriculture, acting as another problematic externality. Consider a society that lives near a forest area. Due to limited space, individuals and businesses will consider expanding into the region and encourage deforestation to cultivate cash crops. In addition, the aforementioned water contamination would affect the abundance of potable water for individuals in proximity. Not only will this provide lower utility to those citizens living in proximity to this area, but it also would drive the house prices down in the area due to the decreased aesthetics of the surroundings, cutting the profits of local housing businesses.

The model for this externality is similar to the model stated in the pollution subsection, where the MPB and MPC intersect with opposing slopes and the MSC is higher than the MPC due to added society cost from the water contamination and aesthetics. However, a solution can be proposed similar to the one in pollution. If the company were to pursue technology businesses that allow for further purification and contaminated less water, increased nitrate content in their used water would not be necessary for their industry; this would make the MSB decrease faster than before (assume other curves follow traditional marginal cost and benefit curves and follow the same law of slopes as pollution cost and benefit curves). Thus, business cooperation can drastically benefit both of the industries, and result in a more socially optimized solution where the externalities of agriculture are reduced or possibly fully eliminated.

## 4. Energy: Industrial Analysis

With an increasing dependency on electricity, fossil fuels, and natural gas, world energy resources play an integral part in global economic stability. Also, with both OECD and non-OECD nations individually contributing to the total primary energy supply, or TPES, production, processing, and trade of energy sources provide valuable information with respect to each country's international economic presence.[14]

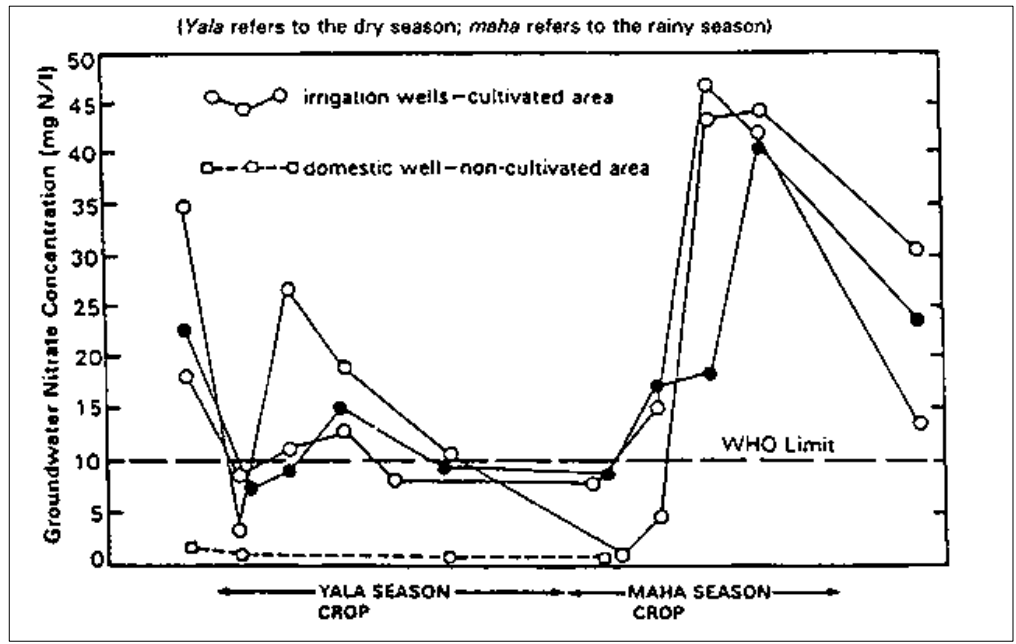

*Figure 3:* A graph of the relationship between the Nitrogen concentration and the month of the year, divided into subsections of dry (Yala) and rainy (Maha) seasons. Important trends include the difference in Nitrate concentrations between irrigation and domestic wells as well as the between the two seasons.[13]

Additionally, energy allocation allows for understanding the relationship between consumption and production, from which additional insight can be gained. Despite OECD countries consuming 53% of the global oil supply, they have a low level of consumption growth. However, although non-OECD countries correspondingly consume 47% of the global supply, they have an increasing rate of consumption.[15] The presented dynamic relationship is just one example of the interdependence between prices, growth, and consumption, all of which are components that can be more effectively presented and understood through an optimized systems model, with specific regard to non-OECD entities.

As non-renewable energy sources quickly deplete, greater reliance falls onto cleaner and more renewable energy sources. Not only does this allow for long-term access to energy, but also eliminates the externality from harmful pollution emitted by the energy sources. However, one needs to keep in mind the distribution of wealth among nations and how unfeasible it is for non-OECD nations to extinguish their resources on technology that permits cleaner energy usage. The charts seen in Figure 4 below represent plots of energy uses stratified by these nations.

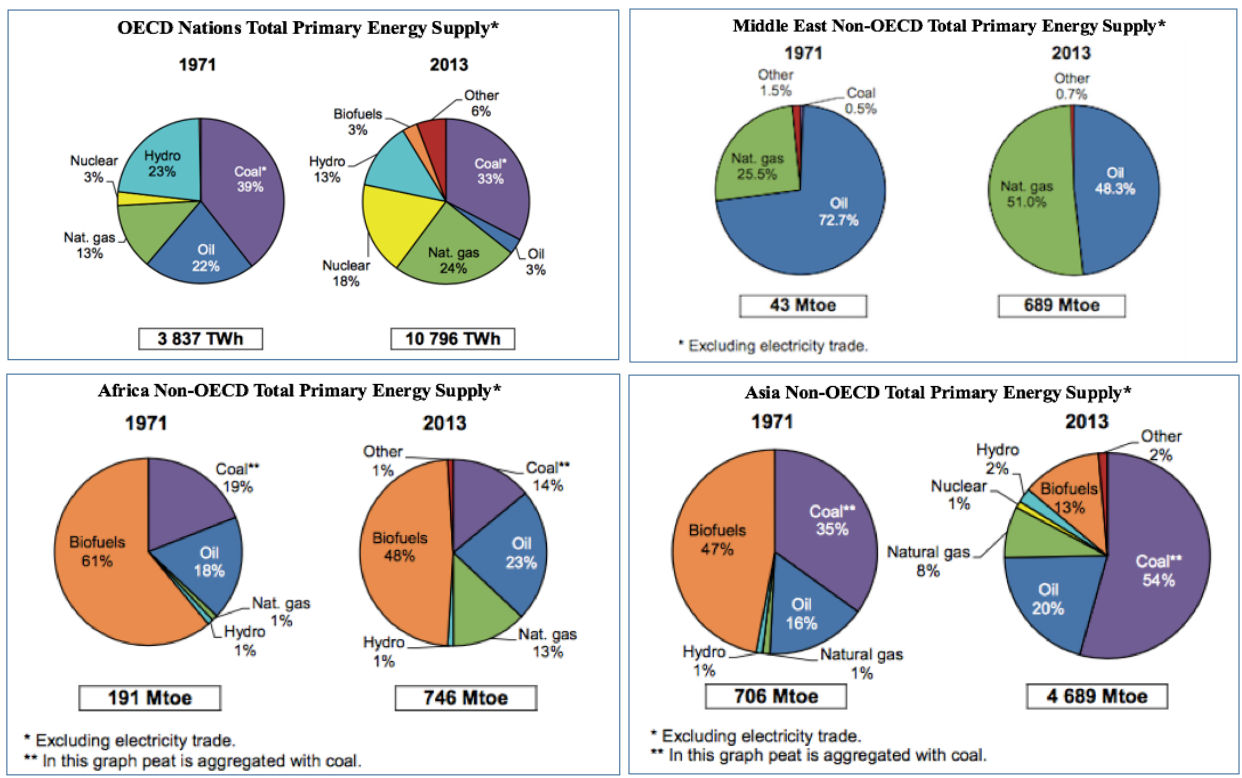

*Figure 4:* The pie charts above, excerpted from the IEA, breaks down the percentage use of the total energy usage of each region into the different energy sources. The first chart combines data from all OECD nations and the other three gather data from Non-OECD nations in Africa, Asia, and the Middle East.[16]

The trends in these charts depict what was previously mentioned about the difference between OECD and non-OECD nations in terms of energy usage. First, we scrutinize the usage distribution of OECD nations in the year 1971. As expected, the total energy usage is only 3837 TWh due to low population as well as low individual usage of energy itself. Additionally, natural gas, oil, and coil (let us name these unclean sources) are the sources that produce the most pollution account for a high 74% of total energy. In contrast, OECD nations have implemented more energy sources, in addition to more efficient and cleaner substitutes. In 2013, the usage of unclean sources has decreased by over 14%, replaced by nuclear power and biofuels, which overlook the minor decrease in hydropower usage. In the other non-OECD nations, we see a reverse trend.

Firstly, analysts may be quick to realize the heavy biofuel usage of Asian, African, and Middle Eastern non-OECD nations in 1971. However, one must note that these nations were mostly in stage two or stage three of the demographic transition model, meaning that they had restricted access to coal and other natural resources; but, in 2013, all three non-OECD regions increased their natural gas and oil usage.[17] Furthermore, it can be assumed that the non-OECD nations are simply at a time lapse of about a quarter decade behind the progress of the OECD nations, which essentially means that the state of non-OECD nations in 2013 was essentially the state of OECD nations in 1971. Thus, the importance of focusing on the key externalities of energy bi-products due to the increased use of such dangerous and non-renewable resources in the non-OECD nations is entirely shown.

One may ask how different types of energy can cause an economic negative externality. Let us consider each type of energy: coal, natural gas, and oil release $SO_2$, $CO_2$, and other $PM_{2.5}$ matter, which as referred to in the pollution subsection, causes major externalities to aesthetics, the environment, and health. Additionally, nuclear power can potentially be dangerous as it emits radioactive material. The cleanest and lowest air pollution emitting materials are hydro energy and biofuels. Hydro-energy simply utilizes the work done by a gravitational field to transfer the gravitational potential energy of the water into kinetic energy and then mechanical energy. The burning of biofuels also sends out little nitrogen content and the remaining matter is absorbed as nutrients by plants. However, non-OECD nations may not have the facilities to be able to use other sources, and with their high CBR rates, require high amounts of energy daily. Thus, we fall back to our initial proposal on business cooperation. Let us discuss the details of this and finally fabricate our optimized system for such externalities.

## 5. Modeling System and Conclusions

Although the previous situations have discussed the theoretical benefits of the model, an optimized model can only be applicable with a thorough mathematical proof. Let us discuss the welfare economics of each situation and proceed in order to highlight the differences. Let the equations:

$$MPC = ax$$
$$MSC = bx$$
$$MSB_1 = -ax + y1$$
$$MSB_2 = cx + y1$$

where $c > b > a$ and $b + c > 2a$. As a result, the socially optimal point for cooperative behavior would occur at the intersection of the MSC and MSB. For non-cooperative behavior, one can equate the functions represented by the MSC and $MSB_1$ to gain the socially optimal point. For cooperative behavior, one can equate the functions represented by the MSC and MSB$_2$. Let O$_1$ and O$_2$, respectively, represent these socially optimal points. Evaluating the difference between the MSC and MPC functions at the x-value of these optimal points results in the value of $\tau_1$ and $\tau_2$, the total amount of Pigouvian taxes needed to correct for the remaining externality and the base of the triangle for $\alpha_1$ and $\alpha_2$. In addition, taking the difference between these x-values and the x-values of the intersection between the MPC and the MSB would provide the height of $\alpha_1$ and $\alpha_2$. The table represented by Figure 5 displays the equation values in terms of a, b, and c.

We can compare the values across cooperative and non-cooperative behavior by looking simply at the denominators:

$$(a + c) > (a + b) \therefore \tau_1 > \tau_2$$
$$2(b + c) > 4a \text{ and } a + c > a + b \therefore \alpha_1 > \alpha_2$$

As a result of cooperative behavior, the amount needed to correct for the externality and fully internalize it is significantly lower than non-cooperative behavior, allowing businesses to allocate the previously lost income to more efficient uses. In addition, it can be noted that the pollution equilibrium point (socially optimal) is lower for cooperative behavior than non-cooperative behavior, so economically the non-OECD nation will mostly reach a lower level of pollution. The area $\alpha$ represents the deadweight loss to the economy and as proved, the total deadweight lost and value of foregone transactions is lower for cooperative behavior than non-cooperative behavior. Conclusively, this microeconomic model allows for optimal business performance and mutual benefit via business cooperation.

An economic analysis has been provided, but a flow chart model would provide another cohesive method to understand these business concepts.

|  | $\tau$ | $\alpha$ |
|---|---|---|
| Non-Cooperative Behavior ($\tau_1$) | $\dfrac{(b-a)y_1}{a+b}$ | $\dfrac{1}{4a}\left[\dfrac{(b-a)y_1}{a+b}\right]^2$ |
| Cooperative Behavior ($\tau_2$) | $\dfrac{(b-a)y_1}{a+c}$ | $\dfrac{1}{2(b+c)}\left[\dfrac{(b-a)y_1}{a+c}\right]^2$ |

*Figure 5:* *The values above utilize the equations of each of the marginal benefit and cost curves for each graph to find the needed Pigouvian tax to correct for the externality and the initial deadweight loss. As seen, comparisons of the denominators can easily show the magnitude of the effect of cooperative behavior on a business.*

Figure 6 depicts a classical model in which the boundaries for perfect cooperation are relayed according to the specific externality situation. This optimized model, along with mathematical substantiation, provides support of emerging business tactics to allow for shared welfare among all sectors and populations of the economy, as well as maximized profits and marginal benefit. The previously overused methods of Pigouvian taxes are hallmarks of inefficiency when other cooperative behavior (not to be mistaken for a collusion) could replace some portion of that amount. Such reduction of taxes would allow for the revenue to be allocated elsewhere and mutually benefit all suppliers and businesses affiliated with the corresponding externality. Additional optimized systems may build upon this hypothesis and provide alternative regression and econometric models to substantiate with both quantitative and qualitative evidence.

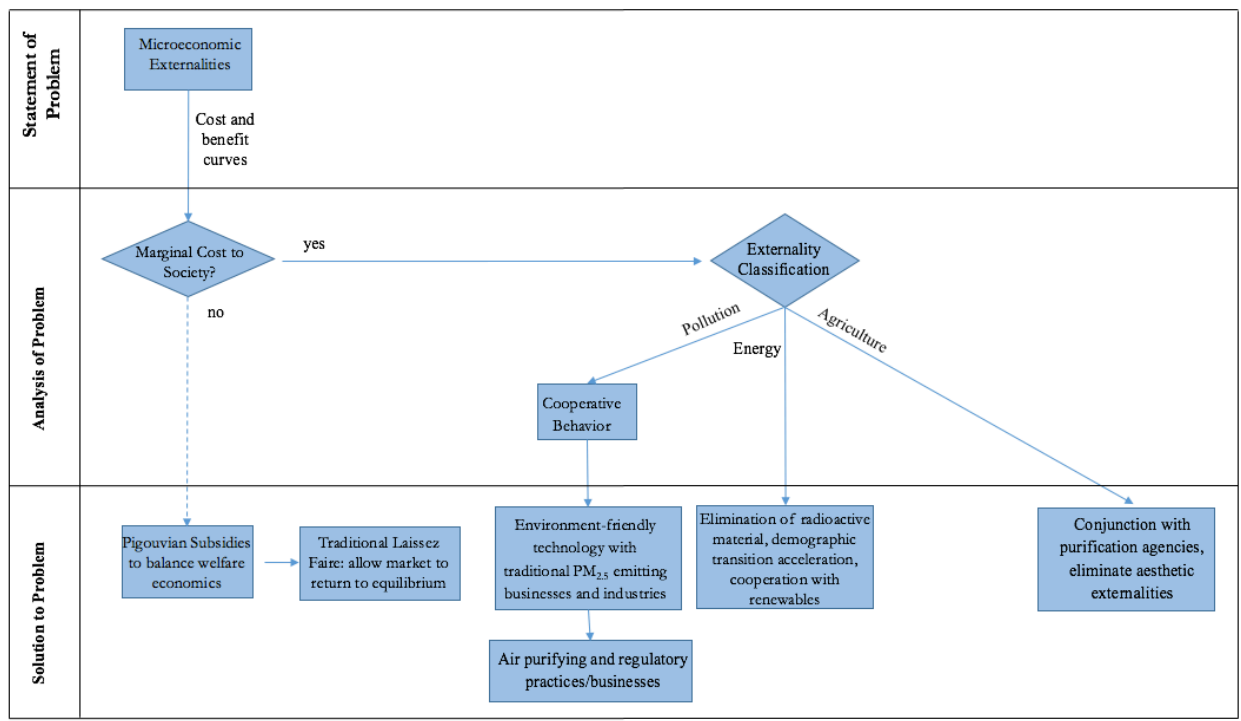

*Figure 6:* This *Systems Analysis Model* summarizes the optimized model for the externality-based solution to the three industrial issues. Broken down into three primary actions and a plan of action/flow chart to follow.